# Ring Artifact and Non-Uniformity Correction Method for Improving XACT Imaging

Mohamed Elsayed Eldib[1]

[1]Department of Radiation Oncology, University of Colorado School of Medicine, Aurora, CO 80045, USA

*Abstract*— X-ray induced acoustic computed tomography (XACT) as a novel imaging modality has shown great potential in applications ranging from biomedical imaging to nondestructive testing. Improve the signal to noise ratio and removing the artifacts are major challenges in XACT imaging. We introduce an efficient non-uniformity correction method for the ultrasound ring transducer. Non-uniformity appears as ring artifacts in the reconstructed image when all sensor elements have a simultaneous time-variant response during the acquisition. Also, non-uniformity causes a contrast change effect in the reconstructed image when some defective sensor elements have a different response than the neighbors over the whole scan time. These two types of non-uniformity appear as horizontal or vertical strip patterns in sinogram domain based on the reason. The proposed method is a sinogram-based-algorithm, which is based on estimating a correction vector and localizing the abnormalities to compensate for the non-uniformity response. We applied the proposed algorithm on both real and simulation data. The results have shown that the proposed method can greatly reduce the ring artifacts and also correct the distortion comes from sensor elements non-uniform response. Despite the great reduction of artifacts, the proposed method does not compromise the original spatial resolution and contrast after using the interpolation. The quantitative analysis has shown a great improvement in the signal-to-noise ratio (SNR) and contrast-to-noise ratio (CNR), while the normalized absolute error (NAE) has been reduced by 77–80 %.

## I. INTRODUCTION

X-RAY induced acoustic computed tomography (XACT) is a promising imaging modality that forms images by detecting the X-ray induced acoustic signal. In XACT, the X-ray is radiated to the object while an ultrasound transducer is used to acquire the generated acoustic wave [1]–[4]. The transducer maybe a single element or array of small elements arranged in a 1D, 2D, cup-shape or ring-shape geometry. There are many applications for the ring ultrasound transducer-based-XACT [5] or photoacoustic tomography (PAT) [6] such as breast cancer screening, where the scanned object is placed in the center of a ring array consists of small elements arranged in a complete or partial circular geometry [7].

However, non-uniform response is a great concern in the ultrasound ring transducer, which degrades the formed image. Non-uniformity may come from scan physical limitations such as crosstalk and reverberation effects during the scan [8], and sensor electronic limitations such as manufacturing issues in the crystals like mis-calibration [9]. There are two types of non-uniformity may occur in the ring array. Firstly, all sensor elements may have a simultaneous time-variant response during the acquisition causing an instantaneous non-uniform response or gain of the signal during the acquisition, which means all sensor elements responses become higher or lower than the normal response at a certain time. This is the most common type and severely degrades the image quality, causing the appearance of ring artifacts in the reconstructed image domain, and stripe patterns in the sinogram domain along sensors positions at a certain time. Secondly, some defective elements may have different responses than the neighbors over the whole scan time causing a non-uniform response or gain of the signal for some sensors during the whole acquisition, which means some sensors responses become higher or lower than the normal neighbor sensors over the whole acquisition time. This type results in changes in contrast components, shadow and streak artifacts in the reconstructed image, and strip artifacts in the sinogram domain along time steps of the acquisition time. XACT is one of state-of-the-art dose imaging techniques in radiological imaging and radiation oncology. So, the resultant change in contrast components of the reconstructed image compromises the dose measurements when using XACT for relative or absolute dosimetry [10], [11].

Low signal-to-noise ratio (SNR) of the X-ray induced acoustic signal (XA signal) further complicates the flat-field correction to reduce the non-uniformity [12]. Even with a flat-field acquisition with a frame averaging, non-uniformity effects often persist in the reconstructed images.

There have been many reports on ring artifact correction methods, some of them can be extended to the ring array-based-XACT. Ring artifact correction can be implemented either on the sinogram domain [9], [13]–[18] or on the reconstructed image domain [19]–[21]. In sinogram, the non-uniformity appears as strip patterns; hence, most sinogram domain approaches estimate the non-uniformity exploiting the feature of stripes in the sinogram, like wavelet-Fourier filtering [16]. In the image domain approaches, the concentric ring components in the reconstructed images are estimated and subtracted from distorted images. Also, ring artifact correction can be performed on both sinogram and reconstructed image sequentially [22].

In this paper, we introduce a sinogram-based-method for ring artifacts correction and sensors non-uniformity correction in the ultrasound ring array. To validate the proposed method, we applied the algorithm on both simulation and experimental data. For simulation, we applied a modified k-Wave-based-simulation on different simulated phantoms to mimic the non-uniform response on the acquired X-ray induced acoustic signal. The results are promising and validate that the proposed algorithm can be generalized for ring artifact correction in other imaging modalities.

## II. MATERIALS AND METHODS

Both ring artifact correction and sensor elements non-uniformity correction are corrected using the same algorithm but the only difference is the direction of correction vector estimation and non-uniformity compensation in the sinogram based on the source of artifacts and non-uniformity, i.e. the direction of stripe patterns in the sinogram domain as shown in Fig. 1. For simplicity, we will discuss the proposed correction method on a ring artifact correction example. Because the main reason of ring artifacts is the instantaneous non-uniform response of the signal during the acquisition for all sensor elements, the non-uniformity response appears as vertical stripe patterns along the sensor positions as shown in Fig. 1(a). We used the k-Wave MATLAB toolbox [23] with a digital phantom to simulate and generate acoustic signal that can be acquired by a virtual ultrasound ring array consists of 360 elements, with a radius of 4.5 mm. The acquired acoustic signal forms the sinogram that was corrupted by multiplying each column of sensors responses at a certain time step by a random factor ranges from 0.1 to 2, which means multiplying the whole original sinogram along sensor positions by a 1D random corruption vector have the same width of the time steps. As a result, stripe patterns appear as non-uniform vertical lines in the sinogram as shown in the distorted sinogram in Fig. 2.

To get the 1D correction vector, as described in Fig. 2, we compute the mean values along the vertical direction in the distorted sinogram to get a 1D mean vector. By taking the inverse of the mean, we get a 1D correction vector. We multiply the correction vector by the distorted sinogram row by row over all sensors positions to get a normalized semi-corrected sinogram free of the stripe patterns but has some loss in contrast components. We can use this advantage to extract the feature of stripes in the sinogram, so we subtract the normalized semi-corrected sinogram from the distorted sinogram to get a difference sinogram that mainly contains the stripe patterns and the contrast change components. Again, we compute the mean values along the vertical direction in the normalized semi-corrected sinogram to get a 1D vector.

To extract the outliers' positions from this mean vector, the outliers are defined as elements more than three local scaled median absolute deviations (MAD) from the local median over a moving window. Empirically, we realized that a widow length of 7 pixels works well in our simulation studies. After determining the positions of outliers, we interpolate the corresponding positions in the distorted sinogram row by row by filling the outliers using a piecewise cubic spline interpolation to get a corrected sinogram free of stripe artifacts without compromising the original spatial resolution or contrast after applying spline interpolation. To reconstruct the images, we applied the 2D iterative time-reversal (ITR) reconstruction on simulation data while filtered-back projection (FBP) reconstruction was applied on the real data. We applied the proposed algorithm on both real data and simulation data corrupted to introduce the two types of non-uniformity, alternatively.

Similarly, same procedures are applied to correct sensor elements non-uniformity by calculating a correction vector along horizontal direction. Entire procedures for ring artifact correction are summarized in Fig. 2. Note that the profiles of outliers' positions and spline fitting are zoomed-in for better visualization.

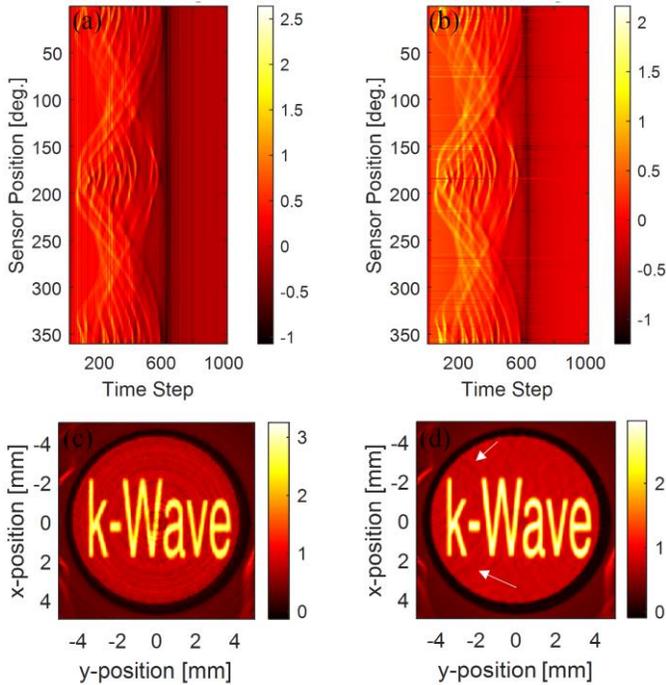

Fig. 1. An example shows the non-uniform responses in the ring array. (a) The corrupted sinogram due to simultaneous time-variant response, (c) the reconstructed image. (b) The corrupted sinogram due to sensor elements non-uniform responses, (d) the reconstructed image.

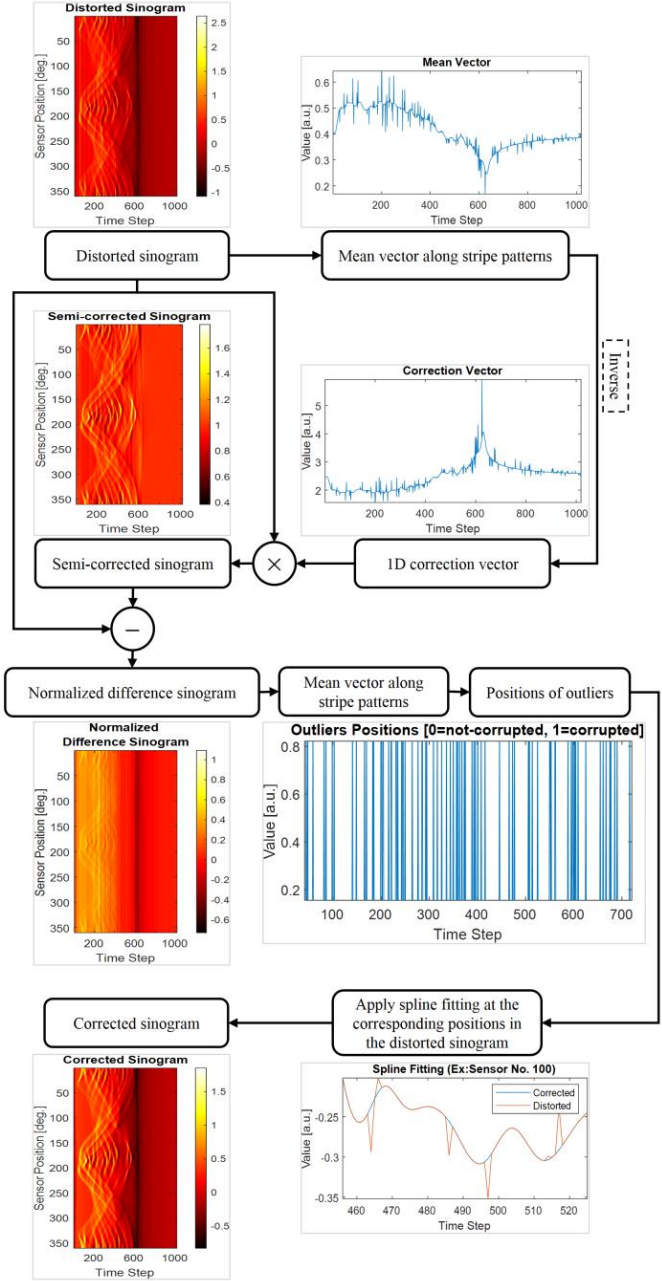

Fig. 2. Flowchart summarizing the main procedures of the proposed method for ring artifact correction.

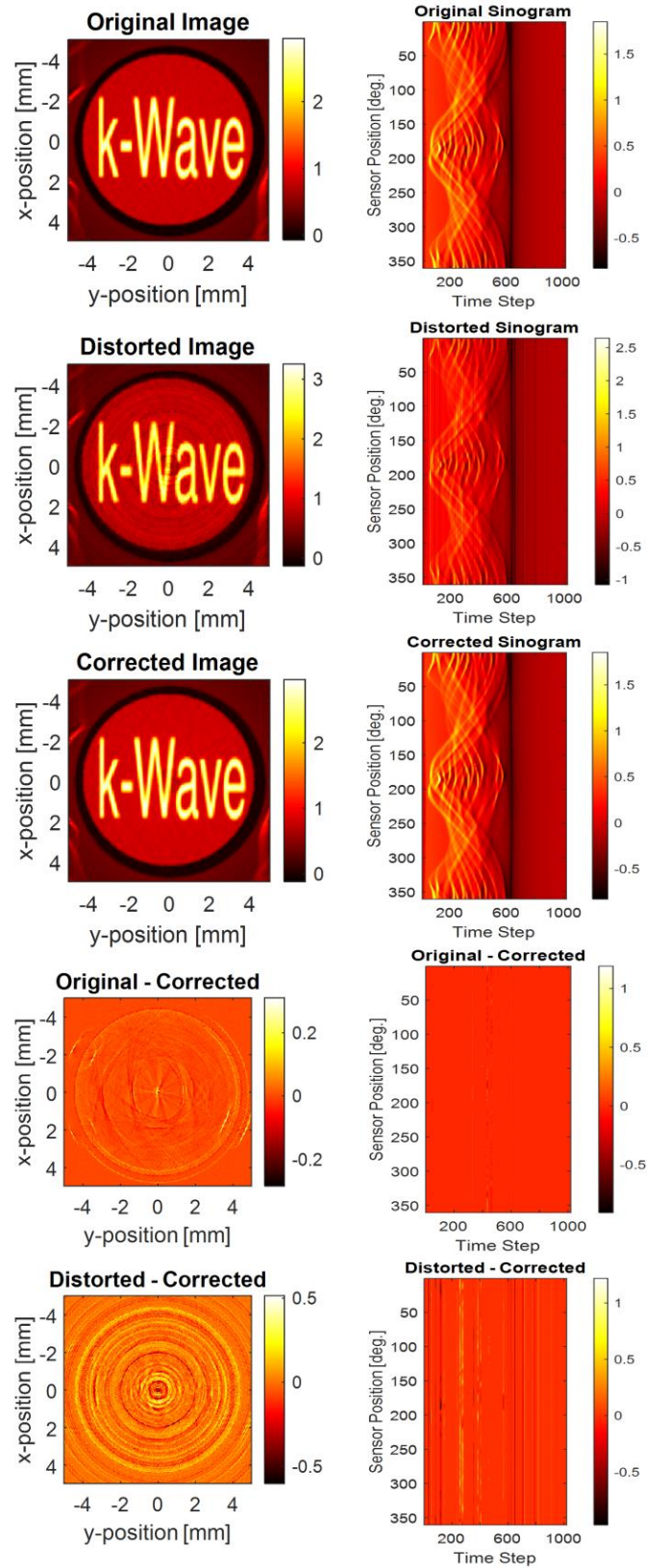

Fig. 3. Simultaneous time-variant response correction in a k-Wave digital phantom shows that vertical stripes in sinogram and ring artifacts in image domain are restored. The left column shows the reconstructed images corresponding to the sinograms in the right column.

## III. RESULTS AND DISCUSSION

The results have shown a great performance of the proposed method for the ring artifacts correction (RAC) and sensor elements non-uniform response correction.

### A. Simultaneous time-variant response correction (RAC)

The proposed algorithm was applied on a k-Wave digital phantom after the formed sinogram was corrupted by multiplying each column of sensors responses at a certain time step by a random factor ranges from 0.1 to 2, which means multiplying the whole original sinogram along sensor positions by a 1D random corruption vector have the same width of the time steps. As a result of corruption, stripe patterns appear as non-uniform vertical lines in the sinogram as shown in the distorted sinogram in Fig. 3, while concentric ring artifacts appear in the reconstructed distorted image. After applying the proposed correction algorithm, the vertical stripe patterns were removed as shown in the corrected sinogram, while the ring artifacts were restored in the corrected image. Neither stripe patterns appear in the difference between the original and corrected sinograms nor ring artifacts appear in the corresponding reconstructed image. The difference between the distorted and corrected sinograms only contains the vertical stripe patterns that caused the ring artifacts, while the corresponding reconstructed image shows the ring components that have been removed.

For more validation on real data, we applied the proposed RAC algorithm on a C-shape real phantom acquired by an ultrasound ring array transducer generating ring artifacts as shown in Fig. 4. In this setup of XACT system, the generated XA signal was acquired by 128 ultrasound transducer elements on the ring array [5]. The imaged sample was placed in the center of the ring array. The result yielded a spatial resolution of 138 μm, from using 5 MHz transducer array. Because of the low SNR of the acquired signal, it was mandatory to get rid of these high frequency nose before applying the proposed method. As a pre-processing step in sinogram domain, we applied wavelet denoising method [24], [25] that consists of bank of filters to remove the high frequency signal that presents the noise in our case, while preserving the low frequency signal that presents the desired signal. Wavelet decomposition parameters have been chosen empirically (Wavelet: db4; Level: 7; Denoising Method: Bayes; Threshold Rule: Median; Noise: Level Independent). The results show a remarkable reduction after applying the proposed RAC algorithm as shown in Fig. 4. The reconstructed images in Fig. 4 represent only the region of interest (ROI) in the entire reconstructed images for better visualization. It is obvious that the ring artifacts severity increases in the middle of black dashed box that indicates the center of the entire image compared to the image peripherals. This validates our assumption that the crosstalk and reverberation can be considered as reasons for ring artifacts.

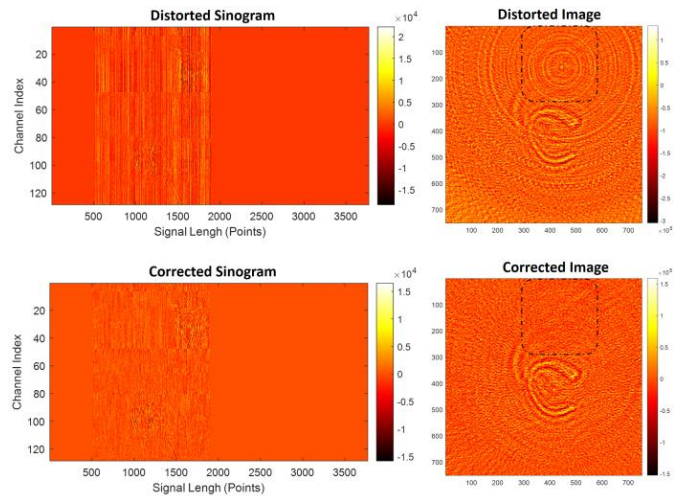

Fig. 4. Ring artifact correction on a C-shape real phantom acquired by an ultrasound ring array transducer.

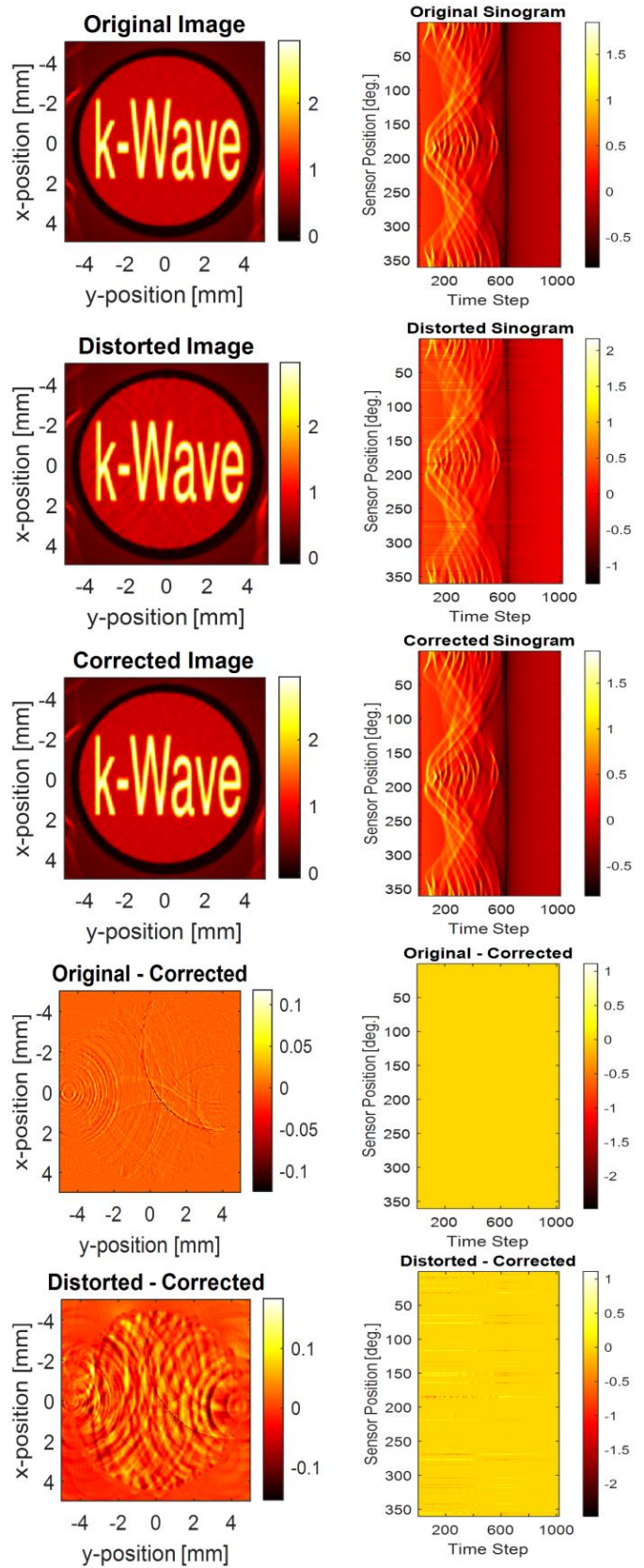

Fig. 5. Sensor elements non-uniformity correction in a k-Wave digital phantom shows that horizontal stripes in sinogram and contrast components in image domain are restored. The left column shows the reconstructed images corresponding to the sinograms in the right column.

## B. Sensor elements non-uniformity correction

The proposed algorithm was applied on a k-Wave digital phantom after the formed sinogram was corrupted by multiplying each row of sensor elements responses over the whole scan time by a random factor ranges from 0.1 to 2, which means multiplying the whole original sinogram along the acquisition time by a 1D random corruption vector have the same width of the number of sensor elements. As a result of corruption, stripe patterns appear as non-uniform horizontal lines in the sinogram as shown in the distorted sinogram in Fig. 5, while contrast component changes, shadow and streak artifacts appear in the reconstructed distorted image. After applying the proposed correction algorithm, the horizontal stripe patterns were removed as shown in the corrected sinogram, while the changes in the contrast components were restored in the corrected image. Neither stripe patterns appear in the difference between the original and corrected sinograms nor residual contrast components appear in the corresponding reconstructed image. The difference between the distorted and corrected sinograms only contains the horizontal stripe patterns that caused the artifacts, while the corresponding reconstructed image contains only the contrast components that have been restored.

## C. Quantitative evaluation

Since we applied the intermediate step of a piecewise cubic spline interpolation to the corresponding distorted positions in the distorted sinogram, the resultant reconstructed image may show some degradation in spatial resolution. To evaluate the spatial resolution after applying the proposed correction algorithm, the ROI indicated by the blue solid rectangle, with a size of $100 \times 60$, in the original image in Fig. 6 was chosen to plot the point spread function (PSF) profiles and the modulation transfer functions (MTF) of the reconstructed images before and after the correction in Fig. 3, and 5 compared to the reference original image. The comparison in Fig. 6 shows a great matching between the original and corrected profiles, while the distorted profile shows a deviation. Fig. 6(a), and (c) show the PSF profiles for the ring artifacts correction and sensor elements non-uniformity correction, respectively, while Fig. 6(b), and (d) show the corresponding MTFs, respectively. The PSF profiles were taken horizontally inside the ROI row by row, and the averaged profile was computed. For better visualization, the PSF profiles have been fitted to smooth curves using the spline interpolation. We computed the MTF curves by taking derivative of the average PSF profile and then taking Fourier transform of the derivatives.

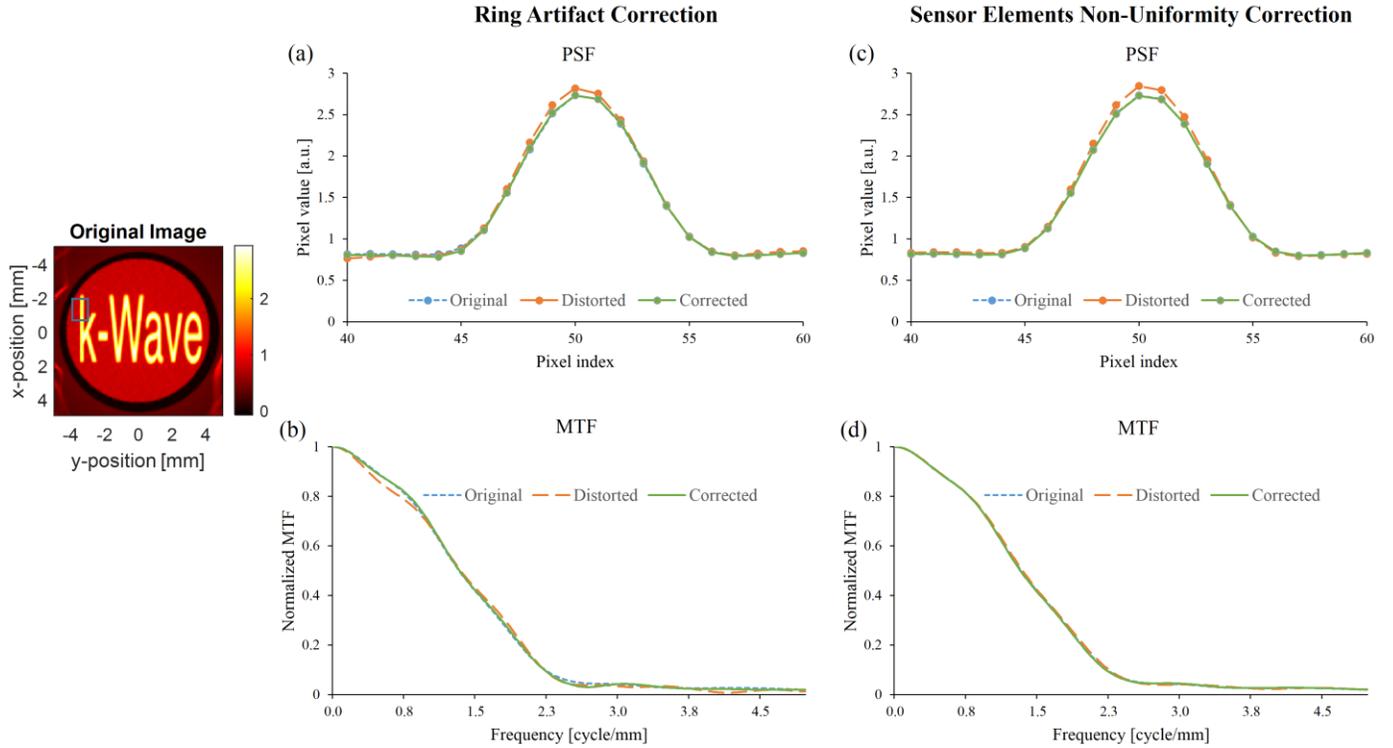

Fig. 6. Quantitative evaluation shows the resolution preservation after correction over the ROI indicated by the solid blue rectangle. (a) PSF corresponding to the same ROI in Fig. 3, (b) the corresponding MTF, (c) PSF corresponding to the same ROI in Fig. 5, (d) the corresponding MTF.

TABLE I

QUANTITATIVE EVALUATION OVER 100 PIXELS OF THE OBJECT AND BACKGROUND IN THE RECONSTRUCTED IMAGES.

| | Ring Artifacts Correction | | | Sensor Elements Non-Uniformity Correction | | |
|---|---|---|---|---|---|---|
| | Original | Distorted | Corrected | Original | Distorted | Corrected |
| **SNR [dB]** | 26.13 | 25.15 | 26.07 | 26.13 | 25.38 | 26.12 |
| **CNR [dB]** | 22.95 | 21.75 | 22.90 | 22.95 | 22.23 | 22.93 |
| **NAE [%]** | 0 | 100 | 21.9 | 0 | 100 | 23.2 |

SNR = signal-to-noise ratio, CNR = contrast-to-noise ratio, NAE = normalized absolute error.

Visual inspection of the sinograms and reconstructed images before and after the correction, shown in Fig. 3, 4, and 5, shows no remarkable difference in terms of spatial resolution. This suggests that the spline interpolation does not compromise the spatial resolution, this is because the piecewise interpolation was applied only to the distorted positions.

For more qualitative evaluation on simulation data, we compared the SNR, contrast-to-noise ratio (CNR), and normalized absolute error (NAE) metrics for the distorted and corrected images corresponding to the reference original images shown in Fig. 3, and 5. We measured the SNRs at the center of the image which shows more ring artifacts. After applying the proposed algorithm, the SNR improved due to the ring artifact reduction and non-uniformity correction. We also measured the CNRs over 100 pixels in the object and background. The CNR also improved due to the correction. The measurements of the NAE represent the error reduction in the corrected image while considering the error in the original image is 0 % and 100 % in the distorted image. The quantitative evaluation shows a great similarity in the SNR and CNR for the original and corrected images, while the NAE has been reduced by ~80 % as summarized in Table 1.

## IV. Conclusion

The proposed method shows a great reduction in ring artifacts and sensor elements non-uniform response. Despite the great reduction of artifacts, the proposed method does not compromise the original spatial resolution or contrast after using the interpolation as an intermediate step. The quantitative analysis has shown a great improvement in the SNR and CNR, while the error NAE has been reduced by ~80 %. We believe that the proposed method can be generalized for more medical imaging modalities like CT imaging.